# Fairness Issues in AI Systems that Augment Sensory Abilities

**Leah Findlater, Steven Goodman**
Human Centered Design & Engineering | DUB Group
University of Washington
Seattle, WA, USA
{leahkf, smgoodmn}@uw.edu

**Yuhang Zhao**
Information Science,
Cornell Tech
Cornell University
New York, NY, USA
yz769@cornell.edu

**Shiri Azenkot, Margot Hanley**
Jacobs Technion-Cornell
Institute, Cornell Tech
Cornell University
New York, NY, USA
shiri.azenkot@cornell.edu

## ABSTRACT
Systems that augment sensory abilities are increasingly employing AI and machine learning (ML) approaches, with applications ranging from object recognition and scene description tools for blind users to sound awareness tools for d/Deaf users. However, unlike many other AI-enabled technologies, these systems provide information that is already available to non-disabled people. In this paper, we discuss unique AI fairness challenges that arise in this context, including accessibility issues with data and models, ethical implications in deciding what sensory information to convey to the user, and privacy concerns both for the primary user and for others.

## Author Keywords
Accessibility; AI fairness; sensory abilities.

## INTRODUCTION
An increasing number of AI-enabled assistive technologies leverage advances in computer vision, speech recognition, auditory scene analysis, and other applications of machine learning to sense and interpret information about the real world for users with disabilities. Active areas of work range from navigation and object recognition tools for blind and low vision users (e.g., [7, 13, 21]) to sound sensing and feedback systems for deaf and hard of hearing users (e.g., [16, 19]).

As the underlying AI advances, these systems will push the boundaries of (dis)ability for people with sensory disabilities. They will ensure that a Deaf person hears a knock on the door or, more fundamentally, shape the way a blind person perceives and engages with people around her. However, unlike many other AI-enabled technologies, these systems provide information that is *already* available to non-disabled people. This goal of equal access raises unique fairness issues for AI-enabled assistive technology: How can we ensure the data and models themselves are accessible? What are the ethical implications of an AI system that determines what information is conveyed to the user about their environment? How can the privacy issues of always-on sensing be mitigated both for the primary user and for others in a larger social sphere?

In this workshop paper, we discuss these questions and outline directions where accessibility and AI/ML researchers can work together to address issues of fairness with AI-enabled assistive technologies for sensory augmentation.

## DATA AND MODEL (IN)ACCESSIBILITY
Since the data used by AI-enabled assistive technologies is inherently not accessible to its primary users, several fairness issues related to data and model accessibility arise. These include questions of how to accessibly explain the system's decision-making process, how to help the user provide effective input to the system, and how to enable the user to verify recognition results.

Observing how AI models interpret input/data is critical for developing appropriate expectations, fostering trust in the system, and making responsible decisions based on its output. With the European Union's "right to explanation" about automated decision-making coming into force in 2018 [10], policy is beginning to reflect a push for transparency in model-based decision-making. Research shows that clarity of an AI model and its limitations can foster trust in the system and drive its continued use [6, 18]. However, in the context of AI-based sensory tools, the data is inherently inaccessible to the target users of the tools, making it difficult for those users to understand the model and thus achieve trust and sound decision-making. For example, one approach to explain an image classification model is to highlight the locations in the image that the model finds to be the most salient or distinguishable (e.g., [18]). These visual explanations, however, are not accessible to blind people, which limits *who* can participate in the development and/or evaluation of the model.

Inaccessible data (and explanations) can also limit the user's ability to personalize their AI system. Kacorri, et al. [12] argue that AI models constrained to an individual user's accessibility needs can work more effectively than more general models. As a case study [13], that same research team explored how visually-impaired people might train personalized object recognition systems with their own snapshots, indicating a need for feedback on the quality of the collected training data (e.g., lighting conditions, number of photos). A challenge of accessible model training, then, is implementing feedback on the quality of collected data in a modality that is accessible to the user. Our sound awareness work for deaf and hard-of-hearing users (e.g., [8]) offers another example, because supporting a d/Deaf user in personalizing their sound classifier requires non-auditory feedback on the quality of each training example. But without the user knowing the characteristics of the sound they are interested in (e.g., possible waveform shape, if the sound is even occurring), how can this feedback be effectively implemented?

Finally, when people with disabilities use AI systems to provide information about their surroundings, inaccurate recognition results can mislead the user, magnifying their vulnerability and even harming their safety (e.g., recognizing a stranger as a friend). This high cost makes it particularly problematic when people with sensory disabilities are not able to determine the accuracy of a system's recognition result. As a consequence, users may place too much trust in the system (e.g., tending to trust inaccurate image description results [14]), or may resort to workarounds that have other negative consequences such as limiting independence (e.g., asking sighted people to double check the system's output) [22]. Thus, one important research question is how to support people with sensory disabilities in independently determining the quality and uncertainty (e.g., confidence level) of AI sensing results.

**DECISION-MAKING IN AI-BASED SENSING**
Because AI-based sensing systems can strongly influence a user's understanding of the world around them, the seemingly simple task of describing sensory input through language is highly complex. In describing a visual or auditory scene, how does a human or AI system decide which details to include and which to leave out?

Consider facial recognition technology for blind and low vision users. Myriad decisions arise in how to implement such a system, with cascading social consequences for the user themselves, and the potential to perpetuate social bias and identify members of marginalized groups without their consent. For example, should such an app only identify by name people the user already knows well, more remote acquaintances, online friends they have never met in person, strangers, celebrities? Besides names, what other information should be conveyed? What about sensitive social categories that sighted people are constantly judging based on visual cues, such as age, gender, and race? On the one hand, an equal access argument can be made that blind users should be able to have this information as well, but this argument is in tension with known issues of bias in AI systems (e.g., [4]). There is also the question of who gets to decide what labels should be used—members of the community being identified, big tech companies, end users?—which reveals the underlying fairness-related tensions at play.

For a less sensitive example, consider a sound awareness system: even in a quiet office, sounds can include a ticking clock, chairs squeaking, typing, a quiet phone conversation, cars and birds outside, the hum of the heating system, and so on. Understandably, a recent study of user needs for sound awareness systems showed that most deaf and hard of hearing participants wanted the system to filter out unimportant sounds, such as ambient background sounds [8]. To what extent can a deaf user trust an AI-based sensing system to notify them of important sounds and to filter out unimportant ones? Who should determine what is an important sound and how that sound is described?

Even this seemingly-mundane use case is loaded with consequences of multiple layers of human decisions, amplified through AI, that shape the way people with sensory disabilities perceive and make sense of the world around them. Who decides what labels to use when labeling training data? Who applies those labels and what biases might they have? Who decides what datasets and modeling approaches should be incorporated into the final system? And how should we present the model's output to the user? Today, we have little visibility into who makes these decisions and how they are made. "Datasheets for Datasets" [9] aims to address this gap by creating prototypes of "datasheets"—forms that accompany every dataset, fostering a level of critical thinking during data collection and transparency afterward. While this is a step in the right direction, it only begins to address the first two layers of decision-making. Researchers should study the entire decision-making pipeline, and consider following Gebru et al.'s approach to improve transparency.

**INDIVIDUAL AND SOCIETAL PRIVACY ISSUES**
Augmenting sensory abilities raises privacy issues both for the primary user of the system and for others, particularly when these AI-based systems employ always-on sensors like wearable cameras or live microphones. As a society we will need to evolve policies and practices that we feel are just yet that also balance the right to privacy.

For the primary user of the system, privacy issues arise from the risk of inadvertently sharing personally sensitive data and from pressure the user may feel to relinquish their privacy in order to achieve the benefits of using the system (e.g., gaining a description of an image or sound clip). A classic example is smartphone-based object recognition for blind users—whether fully AI-driven (e.g., Microsoft SeeingAI [15]) or crowd- and/or human-service- driven (e.g., VizWiz [3], Aira [2]). The recent publication of a VizWiz usage dataset of pictures with private information includes examples ranging from prescription and credit card information to family photos and pregnancy test results [11]. Creators of AI-based sensing systems are responsible for mitigating these privacy risks and for appropriately conveying risks to the user.

Complex privacy concerns also arise when considering the broader social context in which these systems will be used. Conversation partners' and bystanders' attitudes are evolving about the use of always-on, AI-based sensing in public contexts, impacted by whether those individuals realize that the device is providing assistance to the wearer, changing societal norms around wearable technologies in general, and legal and policy implications. For example, Profita et al. [17] showed that third-party onlookers were more accepting of public head-mounted display use (with an onboard camera) when they thought the wearer of the device had a disability, and even more so when they knew the device was being used for an assistive purpose. Ahmed et al. [1] investigated the use of HMDs in a workplace environment, examining what information co-workers were comfortable sharing via these devices. A main takeaway of Ahmed et al.'s study emphasizes the intersection with accessibility: sighted co-workers were more willing to share personal information (e.g., demographics, contact information) with visually impaired colleagues than with other sighted co-workers.

Finally, there are legal and surveillance implications for AI-based assistive sensing systems. Several US states have all-party consent laws, where everyone involved in a conversation must consent to being audio recorded. Lawmakers are also grappling with how to regulate rapidly emerging technologies such as facial recognition software. In 2019, for example, San Francisco banned the use of facial recognition technology by police and other agencies, a response to concerns about the unprecedented opportunities that these technologies offer for government surveillance [5]. For Deaf users, however, audio recording may enable useful sound awareness features, and for blind users, facial recognition software may be useful for getting an understanding of who is around or who is approaching, in both cases providing information that may be seen as equitable to what a non-disabled individual already has access to. How these complexities play out with assistive sensing systems must be carefully considered. One potential path is for there to be "assistive use" exceptions for some of these technologies that are similar to legal exceptions for service animals, where service animals can accompany people with disabilities in all areas that members of the public are normally allowed to go, even places where animals are otherwise not allowed (e.g., the ADA in the US [20]).

## CONCLUSION

We have outlined AI fairness issues that arise with AI-enabled assistive technology for people with sensory disabilities, highlighting open challenges for the accessibility community as well as directions that will be best addressed through close collaboration between AI and accessibility researchers. While the data itself is inherently inaccessible in an assistive sensing context, we can work with AI/ML researchers to address accessibility issues related to model explainability and personalization. Studying the entire decision-making pipeline for AI-based assistive sensing systems is also important, and following Gebru et al.'s [9] approach to improve transparency may be useful here. Finally, as a community, we should establish guidelines to help people reach decisions that balance privacy and fairness considerations for AI sensing, and actively contribute to policy decisions regarding AI, such as the "right to explanation" and privacy protections.

## ACKNOWLEDGMENTS
This work was supported in part by the Office of the Assistant Secretary of Defense for Health Affairs under award W81XWH-14-1-0617 and by the National Science Foundation under award IIS-1763199. We also thank Hal Daumé III for early discussion related to this paper.

## REFERENCES

1. Ahmed, T., Kapadia, A., Potluri, V., & Swaminathan, M. 2018. Up to a limit?: Privacy concerns of bystanders and their willingness to share additional information with visually impaired users of assistive technologies. *Proceedings of the ACM on Interactive, Mobile, Wearable and Ubiquitous Technologies*, 2(3), 89.

2. Aira. 2018. Retrieved July 2, 2019 from https://aira.io/

3. Bigham, J. P., Jayant, C., Ji, H., et al. 2010. VizWiz: nearly real-time answers to visual questions. In *Proceedings of the 23nd Annual ACM Symposium on User Interface Software and Technology*, 333-342. ACM.

4. Buolamwini, J., & Gebru, T. 2018. Gender shades: Intersectional accuracy disparities in commercial gender classification. *Proceedings of the Conference on Fairness, Accountability and Transparency*, 77-91.

5. Conger, K, Fausset, R. and Kovaleski, S.F. 2019, May 14. San Francisco bans facial recognition technology. *New York Times*. Retrieved July 1, 2019 from https://www.nytimes.com/2019/05/14/us/facial-recognition-ban-san-francisco.html

6. Dzindolet, M. T., Peterson, S. A., Pomranky, R. A., Pierce, L. G., & Beck, H. P. 2003. The role of trust in automation reliance. *International Journal of Human-Computer Studies*, 58(6), 697-718.

7. Fiannaca, A., Apostolopoulous, I., & Folmer, E. 2014. Headlock: a wearable navigation aid that helps blind cane users traverse large open spaces. *Proceedings of the 16th international ACM SIGACCESS Conference on Computers & Accessibility*, 19-26. ACM.

8. Findlater, L., Chinh, B., Jain, D., Froehlich, J., Kushalnagar, R., & Lin, A. C. 2019. Deaf and hard-of-hearing individuals' preferences for wearable and mobile sound awareness technologies. *Proceedings of the SIGCHI Conference on Human Factors in Computing Systems (CHI)*, paper 46. ACM.

9. Gebru, T., Morgenstern, J., Vecchione, B., Vaughan, J. W., Wallach, H., Daumé III, H., & Crawford, K. 2018. Datasheets for datasets. arXiv preprint arXiv:1803.09010.
10. Goodman, B., & Flaxman, S. 2017. European Union regulations on algorithmic decision-making and a "right to explanation". *AI Magazine*, 38(3), 50-57.
11. Gurari, D., Li, Q., Lin, C., Zhao, Y., Guo, A., Stangl, A., & Bigham, J. P. 2019. VizWiz-Priv: a dataset for recognizing the presence and purpose of private visual information in images taken by blind people. *Proceedings of the IEEE Conference on Computer Vision and Pattern Recognition*, 939-948.
12. Kacorri, H. 2017. Teachable machines for accessibility. *ACM SIGACCESS Accessibility and Computing*, (119), 10-18.
13. Kacorri, H., Kitani, K. M., Bigham, J. P., & Asakawa, C. 2017. People with visual impairment training personal object recognizers: Feasibility and challenges. *Proceedings of the 2017 CHI Conference on Human Factors in Computing Systems*, 5839-5849. ACM.
14. MacLeod, H., Bennett, C. L., Morris, M. R., & Cutrell, E. 2017. Understanding blind people's experiences with computer-generated captions of social media images. *Proceedings of the 2017 CHI Conference on Human Factors in Computing Systems,* 5988-5999. ACM.
15. Microsoft. (2019). Seeing AI. Retrieved July 2, 2019 from https://www.microsoft.com/en-us/ai/seeing-ai
16. Peng, Y. H., Hsi, M. W., Taele, et al. 2018. SpeechBubbles: Enhancing Captioning Experiences for Deaf and Hard-of-Hearing People in Group Conversations. *Proceedings of the 2018 CHI Conference on Human Factors in Computing Systems*, paper 293. ACM.
17. Profita, H., Albaghli, R., Findlater, L., Jaeger, P., & Kane, S. K. (2016, May). The AT effect: how disability affects the perceived social acceptability of head-mounted display use. *Proceedings of the 2016 CHI Conference on Human Factors in Computing Systems*, 4884-4895. ACM.
18. Ribeiro, M. T., Singh, S., & Guestrin, C. 2016. Why should I trust you?: Explaining the predictions of any classifier. In *Proceedings of the 22nd ACM SIGKDD international conference on knowledge discovery and data mining*, 1135-1144. ACM.
19. Sicong, L., Zimu, Z., Junzhao, D., Longfei, S., Han, J., & Wang, X. 2017. Ubiear: Bringing location-independent sound awareness to the hard-of-hearing people with smartphones. *Proceedings of the ACM on Interactive, Mobile, Wearable and Ubiquitous Technologies*, 1(2), 17.
20. US Department of Justice. 2011. ADA Requirements: 1. Service Animals. Retrieved July 1, 2019 from https://www.ada.gov/service_animals_2010.htm
21. Zhao, Y., Szpiro, S., Knighten, J., & Azenkot, S. 2016. CueSee: exploring visual cues for people with low vision to facilitate a visual search task. *Proceedings of the 2016 ACM International Joint Conference on Pervasive and Ubiquitous Computing*, 73-84. ACM.
22. Zhao, Y., Wu, S., Reynolds, L., & Azenkot, S. 2018. A face recognition application for people with visual impairments: understanding use beyond the lab. *Proceedings of the 2018 CHI Conference on Human Factors in Computing Systems*, paper 215. ACM.